\documentclass[aps,pra,showpacs,groupedaddress,superscriptaddress,twocolumn,showpreprint]{revtex4}

\usepackage[dvips]{graphics,color}

\usepackage{graphicx}

\newcommand{\be}{\begin{equation}}
\newcommand{\ee}{\end{equation}}
\newcommand{\beqn}{\begin{eqnarray}}
\newcommand{\eeqn}{\end{eqnarray}}
\newcommand{\beqnn}{\begin{eqnarray*}}
\newcommand{\eeqnn}{\end{eqnarray*}}

\begin{document}

\title{Microscopic models of quantum jump super-operators}

\author{A. V. Dodonov}
\email{adodonov@df.ufscar.br}

\author{S. S. Mizrahi}
\email{salomon@df.ufscar.br}

\affiliation{Departamento de F\'{\i}sica, CCET, Universidade
Federal de S\~{a}o Carlos, Via Washington Luiz km 235, 13565-905,
S\~ao Carlos, SP, Brazil}

\author{V. V. Dodonov}
\email{vdodonov@fis.unb.br} \affiliation{Instituto de F\'{\i}sica,
Universidade de Bras\'{\i}lia,\\
PO Box 04455, 70910-900, Bras\'{\i}lia, DF, Brazil}

\date{\today}

\begin{abstract}
We discuss the quantum jump operation in an open system, and show
that jump super-operators related to a system under measurement
can be derived from the interaction of that system with a quantum
measurement apparatus. We give two examples for the interaction of
a monochromatic electromagnetic field in a cavity (the system)
with 2-level atoms and with a harmonic oscillator (representing
two different kinds of detectors). We show that derived quantum
jump super-operators have `nonlinear' form $J\rho = \gamma\,
\mbox{diag} \left[ F(\hat{n})a{\rho
}a^{\dagger}F(\hat{n})\right]$, where the concrete form of the
function $F(\hat{n})$ depends on assumptions made about the
interaction between the system and the detector. Under certain
conditions the asymptotical power-law dependence
$F(\hat{n})=(\hat{n}+1)^{-\beta}$ is obtained. A continuous
transition to the standard Srinivas--Davies form of the quantum
jump super-operator (corresponding to $\beta=0$) is shown.

\end{abstract}



\maketitle


\section{Introduction}\label{sec1}


In the theory of continuous photodetection and continuous
measurements the (one-count) quantum jump super-operator (QJS) is
an essential part of the formalism \cite
{carmichael,plenio,ueda1,ueda3,Gard92,WiseMilb,Ban93,Garr94,agarwal,%
ueda2,brun,marsh}, since it accounts for the loss of one photon
from the electromagnetic field (EM) and corresponding
photoelectron detection and counting within the measurement
apparatus (MA). One of the main equations in this theory is the
evolution equation of the field's density operator $\rho _{t}$, or
master equation, which reads in the simplest variant as
\begin{equation}
\frac{d\rho _{t}}{dt}=\frac{1}{i\hbar }\left[ H_{0},\rho _{t}\right] -\frac{%
\gamma }{2}\left( O^{\dagger }O\rho _{t}+\rho _{t}O^{\dagger
}O-2O\rho _{t}O^{\dagger }\right) ,  \label{eqmestra}
\end{equation}
where $H_{0}$ is the EM field Hamiltonian, $\gamma $ is the
field-MA coupling constant and $O$ is some lowering operator,
representing the loss of a single photon from the field to the
environment, that may be detected and counted by a duly
constructed experimental setup. Defining the effective
non-hermitian Hamiltonian as \cite{Mol75,Gisin92,Dum92,Molmer93}
\begin{equation}
H_{eff}=H_{0}-i\frac{\gamma }{2}O^{\dagger }O,  \label{Heff}
\end{equation}
Eq. (\ref{eqmestra}) can be written as (we set here $\hbar =1$)
\begin{equation}
\frac{d\rho _{t}}{dt}= -i \left( H_{eff}\rho _{t}-\rho
_{t}H_{eff}^{\dagger }\right) +\gamma O\rho _{t}O^{\dagger },
\label{eqmestra2}
\end{equation}
whose formal solution is (see, for example,
\cite{carmichael,Zol87})
\begin{eqnarray}
\rho _{t}&=&\sum_{k=0}^{\infty
}\int_{0}^{t}dt_{k}\int_{0}^{t_k}dt_{k-1}\cdot \cdot \cdot
\int_{0}^{t_2}dt_{1}e^{L\left( t-t_{k}\right) }{J}\nonumber
\\
&&\times e^{L\left( t_{k}-t_{k-1}\right) }{J}\cdot \cdot \cdot
{J}e^{Lt_{1}}\rho _{0},
\end{eqnarray}
where
\[
L\rho _{0}=-i\left[ H_{eff}\rho _{0}-\rho _{0}H_{eff}^{\dagger
}\right] ,
\]
$\rho _{0}$ being the density operator for the field state at
$t=0$. The no-count super-operator $\exp \left[ L\left(
t_{k}-t_{k-1}\right) \right] $ evolves the initial state $\rho
_{0}$ from time $t_{k-1}$ to the latter time $t_{k}$ without
taking out any photon from the field, it represents the field
monitoring by a MA. The QJS ${J\bullet }=\gamma O\bullet
O^{\dagger }$ is an operation which takes out instantaneously one
photon from the field. Actually, Tr$\left[ {J}\rho _{0}\right] $
is the rate of photodetection~\cite {SD}.

The explicit form of the QJS is not predetermined. In the
phenomenological photon counting theory developed by Srinivas and
Davies \cite{SD} the QJS was introduced {\em ad hoc} as \be
J_{SD}\bullet=\gamma _{SD}a\bullet a^{\dagger }. \label{JSD} \ee
Later, Ben-Aryeh and Brif~\cite{Aryeh} and Oliveira {\em et
al.}~\cite{Oliveira} considered QJS of the form \be
J_E\bullet=\gamma_E E_-\bullet E_+, \label{JE} \ee where \be
E_-=(a^{\dagger}a+1)^{-1/2}a \quad {\rm and} \quad E_+=E_-^\dagger
\label{E-E+} \ee are the exponential phase operators of Susskind
and Glogower \cite{susg,CarNiet}. These ``non-linear'' operators
allow to remove some inconsistencies of the SD theory noticed by
its authors.

However, the QJS (\ref{JE}) was introduced in \cite{Aryeh,Oliveira}
also {\em ad hoc}. Therefore it is desirable to have not only a
phenomenological theory, but also some {\em microscopic models},
which could justify the phenomenological schemes. The simplest
example of such a model was considered for the first time in
\cite{Imoto}, where the QJS of Srinivas and Davies was derived
under the assumption of highly efficient detection. The two
fundamental assumptions of that model were: (a) infinitesimally
small interaction time between the field and the MA, and (b) the
presence of only few photons in the field mode. Only under these
conditions one can use a simple perturbative approach and arrive
at the mathematical expression for the QJS, which is independent
of the details of interaction between the MA and the EM.

If the conditions (a) or (b) are not fulfilled, the QJS should
depend on many factors, such as, for example, the kind of
interaction between the field and MA, the interaction strength and
the time $T$ of the interaction. Moreover, it should be
emphasized, that the instant $t_{j}$ at which the quantum
jump occurs cannot be determined exactly --- it can happen
randomly at any moment within $T$. Making different assumptions
concerning the moment of `quantum jump', one can obtain different
formal expressions for the QJS. In \cite{AVS} we have proposed a
simple heuristic model for obtaining the `non-linear' QJS of the
form
\be
{J\bullet }=\gamma F(a^{\dagger}a) a\bullet a^{\dagger}
F(a^{\dagger}a).
\label{JF}
\ee
 In this connection, the aim of the present
paper is to provide a more rigorous derivation of QJS\'{}s, using
a more sophisticated model that takes into account dissipation
effects due to the `macroscopic part' of the MA. Our approach is
based on the hypothesis that the transition probability must be
averaged over the interaction time $T$, during which a photon can
be gobbled by the detector at any time in the interval $(0,T) $.
Considering two different models of MA\'{}s: a 2-level atom and a
harmonic oscillator  interacting with a single-mode EM field, we
shall demonstrate that different kinds of interaction result in
quite different QJS\'{}s.

The plan of the paper is as follows. In Sec. II we derive the QJS
using the modified Jaynes--Cummings model (with account of damping
due to the spontaneous decay of the excited state)
and calculating the time
average of the transition operator. In Sec. III we apply the same
scheme to the model of two coupled oscillators, showing explicitly
how the variation of the relative strength of coupling constants
results in the change of the function $F(a^{\dagger}a)$
in Eq. (\ref{JF}).
 Sec. IV contains a summary and conclusions.

\section{Model of two-level atom detector}

\label{2level}

Let us consider first the model, which is a straightforward
generalization of the one studied in \cite{Imoto}. The role of the
`system' is played by a single mode of the electromagnetic field,
while the `detector' of the MA (sub-system constituting the MA
that actually interacts with the EM field) consists of a single
two-level atom. The Hamiltonian for the total system is chosen in
the standard form of the Jaynes--Cummings model \cite{JCM}
\begin{equation}
H_{0}=\frac{1}{2}\omega _{0}{\sigma }_{0}+\omega {\hat{n}}+g{a}{\sigma }%
_{+}+g^{*}{a}^{\dagger }{\sigma }_{-}\,,  \label{ham01}
\end{equation}
where the Pauli pseudo-spin operators ${\sigma }_{0}$ and ${\sigma
}_{\pm }$ correspond to the atom ($\sigma _{+}=|e\rangle \langle
g|$, $\sigma _{-}=|g\rangle \langle e|$ and $\sigma _{0}=|e\rangle
\langle e|-|g\rangle \langle g|$) and one considers that there
were chosen two levels of the atom (the ground state $|g\rangle $
with frequency $\omega _{g}$ and the excited state $|e\rangle $
with frequency $\omega _{e}=\omega _{g}+\omega _{0}$); {$a $,
$a^{\dagger }$ and ${\hat{n}}=a^{\dagger }a$} are the lowering,
rising and number operators, respectively, of the EM field. Since
the coupling between the field and the atom is weak, we assume
that $\omega \gg |g|$. Until now, the detector can absorb and emit
photons back into the EM field, since the detector is not coupled
to some macroscopic device that irreversibly absorbs the photons.

Therefore, we have to take into consideration that the detector is
coupled to the `macroscopic part' (MP) of the MA (e.g., phototube
and associated electronics). Hence the detector suffers
dissipative effects responsible for the spontaneous decay of the
excited level of the detector (in this case of the atom). And it
is precisely this physical process that represents a
photodetection -- the excited level of the detector decays,
emitting a photoelectron into the MP of the MA, which is amplified
by appropriate electronics and is seen as a macroscopic electrical
current inside the MP of MA. We can take into account this
dissipation effects by describing the whole photodetection
process, including the spontaneous decay, by the master equation
\begin{equation}
\frac{d{\rho }_{t}}{dt}+i\left( {H}_{eff}{\rho }_{t}-{\rho }_{t}{H}%
_{eff}^{\dagger }\right) =2\lambda {\sigma }_{-}{\rho }_{t}{\sigma
}_{+}, \label{mestra1}
\end{equation}
which is the special case of Eq. (\ref{eqmestra2}), where $O={\sigma
}_{-}$, $O^{\dagger }={\sigma }_{+}$, ${H}_{eff}=H_{0}-i\lambda
{\sigma }_{+}{\sigma }_{-}$, and $2\lambda $ is the coupling of the
excited level of the atom
(detector) to the MP of the MA (here we make a reasonable assumption that $%
\lambda $ has the same order of magnitude as $|g|$). The `sink'
term
\begin{equation}
{R}\bullet = 2\lambda {\sigma }_{-}\bullet {\sigma }_{+}
\end{equation}
represents the $|e\rangle \rightarrow |g\rangle $ transition
within the detector (the atomic decay process in this case). If
$\lambda=0$, then the detector interacts with the EM field, but
photoelectrons are not emitted (thus no counts happen), because
the absorbed photons are emitted back to the field and then
reabsorbed at a later time, periodically, analogously to the Rabi
oscillations.

In the following, we shall use the quantum trajectories approach
\cite{carmichael}. The effective Hamiltonian (\ref{Heff}) becomes
\begin{eqnarray}
{H}_{eff} &=& {H}-i{\lambda }{\sigma }_{+}{\sigma }_{-}=\frac{1}{2}%
\left( \omega _{0}-i{\lambda }\right) {\sigma }_{0} \nonumber \\
&& +\omega \hat{n}+g{a}{\sigma }_{+}+g^{*}{a}^{\dagger }{\sigma
}_{-} -i\lambda/2 \label{efet} \end{eqnarray} (where we have used
$\sigma _{+}\sigma _{-}=(1+\sigma _{0})/2$) and the evolution of the
system between two spontaneous decays is given by the no-count
super-operator
\begin{equation}
{\cal D}_{t}{\rho }_{0}={U}(t){\rho }_{0}{U}^{\dagger }(t), \qquad
{U}(t)=\exp \left( -i{H}_{eff}t\right) .
\end{equation}
After a standard algebraic manipulation \cite{JCM,Cress96} we
obtain the following explicit form of the {\em non-unitary}
evolution operator $U(t)$:
\begin{eqnarray}
{U}(t) &=&e^{-\lambda t/2}\exp \left[ -i\omega \left( {\sigma
}_{0}/2+{\hat{n}}\right) t\right] \nonumber   \\ &&\times \left\{
\frac{1}{2}\left[ C_{\hat{n}+1}(t) -i\frac{\delta }{\left|
g\right| } S_{\hat{n}+1}(t)
\right] \left( 1+\sigma_{0}\right) \right.   \nonumber \\
&&\left. -i\frac{g}{\left| g\right| } S_{\hat{n}+1}(t) {a}{\sigma
}_{+}-i\frac{g^{*}}{\left|g\right| } {a}^{\dagger }
S_{\hat{n}+1}(t) {\sigma }_{-}
\right.   \nonumber \\
&&\left. +\frac{1}{2}\left[ C_{\hat{n}}(t) +i\frac{\delta }{\left|
g\right| } S_{\hat{n}}(t) \right] \left( 1-\sigma _{0}\right)
\right\} ,
\label{aray}
\end{eqnarray}
where \begin{equation}C_{\hat{n}}(t) \equiv \cos \left(
\left|g\right| {B_{\hat{n}}}t\right) , \quad S_{\hat{n}}(t) \equiv
\sin \left( \left|g\right| {B_{\hat{n}}}t\right) /{B}_{\hat{n}},
\label{def-CS}
\end{equation}
\begin{equation}
{B}_{\hat{n}}=\sqrt{{\hat{n}} + \left({\delta }/{|g|}\right)
^{2}},\qquad \delta =\frac{1}{2}\left( \omega _{0}-\omega
-i\lambda  \right) \label{def-Bd}
\end{equation}
(note that parameter $\delta $ is complex and $\hat{n}$ is an
operator).

Assuming that the field state is ${\rho }_{0}={\rho }%
_{F}\otimes |g\rangle \langle g|$ at time $t=0$
 or, analogously, the last
photoemission occurred at $t=0$, the probability that the {\em next}
photoelectron emission will occur within the time interval
$[t,t+\Delta t)$ is given by \cite{carmichael,SD,Cohen}
\begin{equation}
P(t)={\rm Tr}_{F-D}\left[ R{\cal {D}}_{t}{\rho }_{0}\right] \Delta
t, \label{p}
\end{equation}
(the subscripts $F$ and $D$ are a reminder that the trace
operation is on {\em field} and {\em detector} spaces,
respectively) where $\Delta t$ is the time resolution of the MA.
Tracing out first over the detector variables, the probability
density for the next photoemission to occur at
time $t$ will be~\cite{Cohen} 
\begin{equation}
p(t)=\lim_{\Delta t\rightarrow 0}\frac{P(t)}{\Delta t}={\rm Tr}%
_{F}\left[ \Xi (t){\rho }_{F}\right] ,  \label{salto1}
\end{equation}
where the time-dependent {\em transition super-operator}
\begin{equation}
{\Xi }(t)\bullet =2\lambda {\Gamma }(t)\bullet {\Gamma }^{\dagger
}(t), \label{xi}
\end{equation}
acting on the EM field, stands for the photoelectron emission into
the MP of the MA (i.e., the actual photodetection). Once again,
the probability for detecting a photoelectron in $[t,t+\Delta t)$
is $P(t)={\rm Tr}_{F}\left[
\Xi (t)\rho _{F}\right] \Delta t$ (now on we omit the subscript and write $%
\rho _{F}\equiv \rho $ for the field operator). In Eq. (\ref{xi})
$\Gamma (t) $ is the time-dependent {\em transition operator}
\begin{equation}
{\Gamma }(t)=\langle e|{U}(t)|g\rangle ,  \label{gamma}
\end{equation}
that takes out a single photon from the field state. Substituting
Eq. (\ref{aray}) into Eq. (\ref{gamma}) we can write ${\Gamma
}(t)$ as
\begin{equation}
{\Gamma }(t)=-i\frac{g}{|g|}\exp\left(-\lambda t/2-i\omega
{\hat{n}}t\right) S_{\hat{n}+1}(t){a},
\end{equation}
so the time-dependent transition super-operator (\ref{xi}) becomes
\begin{equation}
{\Xi }(t){\rho }=2\lambda e^{-\lambda t}e^{-i\omega {\hat{n}}t}
S_{\hat{n}+1}(t){a}{\rho }{a}^{\dagger} S_{\hat{n}+1}^{\dagger
}(t) e^{i\omega {\hat{n}}t}. \label{expr}
\end{equation}
In the resonant case, $\omega _{0}=\omega $, we have \be
B_{\hat{n}}=\sqrt{\hat{n}- \chi^{2}}, \qquad \chi \equiv \lambda
/(2|g|). \label{Bn-chi} \ee

If the interaction time $\Delta t$ is small, and the number of
photons in the field is not very high, in the sense that the
condition
\begin{equation}
|g|\Delta t\sqrt{n+1}\ll 1  \label{important}
\end{equation}
is fulfilled for all eigenvalues of $\hat{n}$, for which the probabilities $%
p_{n}=\langle n|\rho |n\rangle $ are important, then one can
replace the operator $\sin \left( B_{\hat{n}+1}|g|\Delta t\right)
$ in Eq. (\ref{def-CS}) simply by $B_{\hat{n}+1}|g|\Delta t$ and
arrive at the QJS
\begin{equation}
{J}{\rho }=e^{-i\omega {\hat{n}}\Delta t}\left[2 \lambda \left(
|g|\Delta t\right) ^{2}{a}{\rho }{a}^{\dagger }\right] e^{i\omega
{\hat{n}}\Delta t}, \label{almost-SD}
\end{equation}
which has {\em almost\/} the Srinivas--Davies form (\ref{JSD}),
with the coupling constant
\begin{equation}
\gamma _{SD}=2\lambda \left( |g|\Delta t\right) ^{2}.
\label{gamasd}
\end{equation}
Taking $2\lambda = (\Delta t)^{-1}$ we obtain the same coupling
constant $\gamma _{SD}$ as in \cite{Imoto}, but this assumption is
not the only possible.
Note that super-operator (\ref{almost-SD}) contains the factors $%
\exp (\pm i\omega {\hat{n}}\Delta t)$, which can be essentially
different from the unit operator even under condition (\ref
{important}), for two reasons: (1) the condition $|g|\Delta t\ll
1$ does not
imply $\omega \Delta t\ll 1$, because $\omega \gg |g|$; (2) the condition (%
\ref{important}) contains the square root of $n$, whereas the
eigenvalues of $\exp (\pm i\omega {\hat{n}}\Delta t)$ depend on the
number $n$ itself, which is much greater than $\sqrt{n}$ if $n\gg
1$. Consequently, even the simplest microscopic model gives rise
to a QJS, which is, strictly speaking, different from the SD jump
super-operator, coinciding with the former only for the diagonal
elements $\left| n\right\rangle \left\langle n\right| $ of the
density matrix in the Fock basis.

If condition (\ref{important}) is not satisfied, we propose that
the QJS can be defined by {\em averaging\/} the transition
super-operator (\ref{expr}) over the interaction time $T$, because
the exact instant within $(0,T)$ at which the photodetection
occurs in each run is unknown, so a reasonable hypothesis is that
these events happen randomly with uniform probability
distribution:
\begin{equation}
{J}_{T}{\rho }=\frac{1}{T}\int_{0}^{T}dt\ {\Xi }(t){\rho }.
\label{J-Xi}
\end{equation}
Writing the field density operator as
\begin{equation}
{\rho }=\sum_{m,n=0}^{\infty }\rho _{mn}|m\rangle \langle n|,
\label{dens2}
\end{equation}
we have
\begin{equation}
{J}_{T}{\rho }= \sum_{m,n=1}^{\infty }\rho _{mn}
\sqrt{mn} f_{mn}
|m-1\rangle \langle n-1|, \label{rho-f}
\end{equation}
where
\begin{equation}
f_{mn} = \frac{2\lambda}{T}
\int_{0}^{T}e^{i\omega t(n-m) -\lambda t }
S_{m}(t)S_{n}(t) \, dt.
\label{fmn}
\end{equation}

It is natural to suppose that the product $\lambda T$ is big
enough, so that the photodetection can happen with high
probability. Mathematically, it means that we assume that
$\exp(-\lambda T) \ll 1$. If $\lambda \ll \omega$ (this is also a
natural assumption), then the off-diagonal coefficients $f_{mn}$
with $m\neq n$ are very small due to fast oscillations of the
integrand in Eq. (\ref{J-Xi}), so they can be neglected (a rough
estimation gives for these terms the order of magnitude ${\cal
O}(\lambda/\omega)$, compared with the diagonal coefficients
$f_{nn}$). Consequently, the microscopic model leads to the
nonlinear {\em diagonal\/}  QJS of the form
 \begin{equation}
 J\rho = \gamma\, \mbox{diag}
\left[ F(\hat{n})a{\rho }a^{\dagger}F(\hat{n})\right],
\label{J-diag}
\end{equation}
where $\mbox{diag}(\hat{A})$ means the diagonal part of the
operator $\hat{A}$ in the Fock basis. The function $F({n})$ can be
restored from the coefficients $f_{nn}$ (apart the constant factor
which can be included in the coefficient $\gamma$) as \be F({n})
=\sqrt{f_{n+1,n+1}}. \label{F-f} \ee

Under the condition $\exp(-\lambda T) \ll 1$, the upper limit of
integration in Eq. (\ref{J-Xi}) can be extended formally to
infinity, with exponentially small error. Then, taking into
account the definition of the function $S_n(t)$ (\ref{def-CS}), we
arrive at integrals of the form
\[
\int_0^{\infty} dt\, e^{-\lambda t}\times
\left\{
\begin{array}{ll}
\sin^2(\mu t)/\mu^2 & {\rm for} \;\; \chi<1 \\
t^2 & {\rm for} \;\; \chi=1 \\
 \sinh^2(\mu t)/\mu^2 & {\rm for} \;\; \chi>1
\end{array}
\right. ,
\]
which can be calculated exactly (see, e.g., Eqs. 3.893.2 and
3.541.1 from \cite{Grad}). The final result does not depend on
$\lambda$ or $\chi$ (and it is the same for either $\chi<1$ or
$\chi>1$): \be f_{nn} = (nT)^{-1}. \label{fnT} \ee

Thus we obtain the QJS
\begin{equation}
{J}_{T}\rho ={\gamma }_{T} \sum_{n=1}^{\infty }\rho _{nn}
|n-1\rangle \langle n-1| ={\gamma }_{T}\mbox{diag}\left(
{E}_{-}{\rho }{E}_{+}\right), \label{JT}
\end{equation}
where ${\gamma }_{T}=T^{-1}$,
and the operators $E_-$ and $E_+$ are defined by Eq.
(\ref{E-E+}).
 Notice that, in principle, ${\gamma }_{T}$
{\em is different} from $\gamma _{SD}$. Moreover, the
super-operator (\ref{JT}) derived from the microscopic model turns
out to be {\em different\/} from the phenomenological QJS
(\ref{JE}) studied in \cite{Oliveira,AVS}. The difference is that
${J}_{T}$ has no off-diagonal matrix elements, while ${J}_{E}$
has. We see that the microscopic model concerned (which can be
justified in the case of big number of photons in the field mode)
predicts that each photocount not only diminishes the number of
photons in the mode exactly by one, but also destroys off-diagonal
elements, which means the total decoherence of the field due to
the interaction with MA.

Note, however, that the formula (\ref{fnT}) holds under the
assumption that the upper limit of integration in Eq. (\ref{fmn})
can be extended to the infinity. But this cannot be done if
parameter $\chi$ is very big. Indeed, for $\chi>1$ and $\lambda T
\gg 1$, the integrand in (\ref{fmn}) at $t=T$ is proportional to
$\exp\left[-\lambda T\left(1- \sqrt{1-n/\chi^2}\right)\right]$, so
it is not small when $n/\chi^2\ll 1$. Calculating the integral in
the finite limits under the conditions $n/\chi^2\ll 1$ and
$\lambda T \gg 1$, we obtain the approximate formula \be
f_{nn}=(Tn)^{-1}\left\{1-\exp\left[-\lambda
Tn/\left(2\chi^2\right) \right]\right\}, \label{interp} \ee which
shows that $f_{nn}$ does not depend on $n$ if $\lambda
Tn/\left(2\chi^2\right) \ll 1$. Thus we see how the QJS (\ref{JT})
can be
 continuously transformed to the SD jump
super-operator (\ref{JSD}), when the number $n$ changes from big
to relatively small values. It should be emphasized, nonetheless,
that the off-diagonal coefficients $f_{mn}$ remain small even in
this limit. Their magnitude approaches that of the diagonal
coefficients only in the case of $\lambda \sim \omega$, which does
not seem to be very physical.

\section{Model of harmonic oscillator detector}

Now let us consider another model, where the role of the detector
is played by a harmonic oscillator interacting with one EM field
mode. This is a simplified version of the model proposed by
Mollow~\cite{mollow} (for its applications in other areas see,
e.g., \cite{DK96} and references therein). In the rotating wave
approximation (whose validity was studied, e.g., in Ref.
\cite{Estes68}) the Hamiltonian is
\begin{equation}
{H}=\omega _{a}{a}^{\dagger }{a}+\omega _{b}{b}^{\dagger }{b}+g{a}{b}%
^{\dagger }+g^{*}{a}^{\dagger }{b},  \label{ham1}
\end{equation}
where the mode ${b}$ assumes the role of the detector and the mode
${a}$ corresponds to the EM field ($\omega _{b}$ and $\omega _{a}$
are the corresponding frequencies and $g$ is the detector-field
coupling constant). In the following we shall repeat the same
procedures we did in the section \ref{2level}. The dissipation
effects due to the macroscopic part of the MA, associated to the
mode $b$, can be taken into account by means of the master
equation in the form
\begin{equation}
\frac{d{\rho }}{dt}+i\left[ {H}_{eff}{\rho }
- \rho {H}_{eff}^{\dagger}\right] =2\lambda {b}{\rho }{b}%
^{\dagger }.  \label{mestra2}
\end{equation}
with the effective Hamiltonian
\begin{eqnarray}
{H}_{eff}&=&{H}-i{\lambda }{b}^{\dagger }{b}=\left( \omega _{b}
-i \lambda \right) {b}^{\dagger }{b}\nonumber \\
&&+\omega _{a}{a}^{\dagger }{a}+g{b}{a}%
^{\dagger }+g^{*}{b}^{\dagger }{a}. \label{quadra}
\end{eqnarray}

The evolution operator $U(t)=\exp (-iH_{eff}t)$ for the {\em
quadratic\/} Hamiltonian (\ref{quadra}) can be calculated by means
of several different approaches \cite{Dbook}. Here we use the
algebraic approach \cite {Ban93,Wei,alg,inter}, since Hamiltonian
(\ref{quadra}) is a linear combination of the generators of
algebra $su(1,1)$
\[
{K}_{+}\equiv {b}^{\dagger }{a},\quad {K}_{-}\equiv
-{b}{a}^{\dagger },\quad {K}_{0}\equiv ({b}^{\dagger
}{b}-{a}^{\dagger }{a})/2,
\]
\[
\lbrack {K}_{0},\ {K}_{\pm }]=\pm {K}_{\pm },\quad [{K}_{-},\ {K}_{+}]=2{K}%
_{0}.
\]
The evolution operator can be factorized as
\begin{equation}
{U}(t)=e^{-i\Omega
t{N}}e^{A(t){K}_{+}}e^{B(t){K}_{0}}e^{C(t){K}_{-}}, \label{evol1}
\end{equation}
where
\[
{N}\equiv \left( {b}^{\dagger }{b}+{a}^{\dagger }{a}\right)
/2,\quad \Omega \equiv \omega _{b}+\omega _{a}-i\lambda .
\]
The time-dependent coefficients are
\begin{equation}
A(t)= -\frac{ig^{*}\sin (\eta t)}{\eta \Upsilon \left( t\right) },
\quad
C(t)= \frac{ig\sin (\eta t)}{\eta \Upsilon \left( t\right) },
\label{AC}
\end{equation}
\be
B(t)=-2\ln \Upsilon \left( t\right),
\label{B}
\ee
with
\begin{equation}
\Upsilon \left( t\right) =\cos (\eta t) + i\left[\omega_{ba}/(2\eta)
\right] \sin(\eta t), \label{def-Ups}
\end{equation}
\begin{equation}
\omega _{ba}\equiv \omega _{b}-\omega _{a}-i\lambda , \quad \eta
\equiv \left( |g|^{2}+\omega _{ba}^{2}/4\right) ^{1/2}.
\label{eqs2}
\end{equation}
Assuming that the detector is in resonance with the EM field's
mode one gets $\omega _{ba}=-i\lambda$ and
\begin{equation}
\Upsilon \left( t\right) =\cos (\eta_0 t) + \left[\lambda/(2\eta_0)
\right] \sin(\eta_0 t), \label{eqs2a}
\end{equation}
\be
\eta_0 = \left( |g|^{2} -\lambda^{2}/4\right) ^{1/2}.
\label{eta0}
\ee
If, initially, the detector oscillator is in the ground state
$|0_{b}\rangle $, the time-dependent transition operator,
corresponding to the absorption of one photon from the EM field,
defined in (\ref{gamma}), is
\begin{eqnarray}
{\Gamma }(t)&=&\langle 1_{b}|U(t)|0_{b}\rangle \nonumber
\\
&=&A(t)\exp \left[ -\frac12\left( i\Omega t+B(t)\right) ({a}^{\dagger
}{a}+1)\right] {a} \label{G2}
\end{eqnarray}
and the transition super-operator becomes
\begin{eqnarray}
\Xi (t)\rho &=& 2\lambda |A(t)|^{2}\exp \left[ -\frac12\left( i\Omega
t+B(t)\right) (a^{\dagger }a+1)\right]
\nonumber \\ && \times
a\rho a^{\dagger }
\exp \left[ \frac12\left( i\Omega^{*}t
-B^{*}(t)\right) (a^{\dagger }a+1)\right] .
\end{eqnarray}

For ``small'' $t=\Delta t$ and few photons in the cavity, the QJS
(\ref{almost-SD}) is recovered. Considering, instead, the
time-averaged QJS, one has Eqs. (\ref{J-Xi})-(\ref{rho-f}). For
$\chi =\lambda/(2|g|)<1$ (when the parameter $\eta_0$ is real) one
can represent the coefficients $f_{mn}$ as (we consider the
resonance case with $\omega_a=\omega_b =\omega$)
\begin{eqnarray}
f_{mn} &=& \frac{4\chi }{T(1-\chi^2)^{3/2}}
 \int_{0}^{Z} dz\, [\cos( z) +\xi \sin(z)]^{m+n-2}
 \nonumber \\ && \times
 \sin^2( z)
 \exp\left[i\overline\omega z(n-m)-\xi z(m+n) \right],
\label{fmn-osc}
\end{eqnarray}
where \be \xi=\frac{\chi}{\sqrt{1-\chi^2}}, \quad \overline\omega=
\frac{\omega}{|g|\sqrt{1-\chi^2}}, \quad Z = \frac{\lambda
T}{2\xi}. \ee Since the parameter $\overline\omega$ is big, the
off-diagonal coefficients $f_{mn}$ with $n \neq m$ are very small
due to the strongly oscillating factor $\exp[i\overline\omega
t(n-m)]$. Consequently, they can be neglected in the first
approximation, and we arrive again at the diagonal QJS of the form
(\ref{J-diag}).

We notice that the exact analytical expression for the integral in
Eq. (\ref{fmn-osc})  is so complicated (even if $m=n$), that it is
difficult to use it. For example, in the limit $\chi \to 1$ Eq.
(\ref{fmn-osc}) can be reduced to the form \be f_{nn} =\frac{4}{T}
\int_{0}^{\lambda T/2} dy\, y^2(1+y)^{2n-2} \exp(-2ny).
\label{int-1} \ee Replacing the upper limit by the infinity, we
recognize the integral representation of the Tricomi confluent
hypergeometric function $\Psi(a;c;z)$ \cite{Bate}. Thus we have
(neglecting small corrections of the order of $\exp(-\lambda T)$)
\be f_{nn}=\frac{8}{T}\Psi(3;2n+2;2n). \label{1-Psi} \ee Although
the $\Psi$-function in the right-hand side of Eq. (\ref{1-Psi})
can be rewritten in terms of the associated Laguerre polynomials
\cite{Bate} as \be \Psi(3;2n+2;2n)= \frac{(2n)!}{2(2n)^{1+2n}}
L_{2n-2}^{(-1-2n)}(2n), \label{Psi-Lag} \ee neither Eq.
(\ref{1-Psi}) nor Eq. (\ref{Psi-Lag}) help us to understand the
behavior of the coefficient $f_{nn}$ as function of $n$. Therefore
it is worth trying to find simple approximate formulas for the
integral in (\ref{fmn-osc}).

If $\chi\ll 1$, then also $\xi \ll 1$, so we can neglect the term
$\xi\sin(z)$ in the integrand of Eq. (\ref{fmn-osc}) and the
function $\sin^2( z)[\cos( z)]^{2n-2}$ can be replaced by its
average value taken over the period $2\pi$ of fast (in the scale
determined by the characteristic time $\xi^{-1}$) oscillations.
After simple algebra we obtain (replacing the upper limit of
integration $Z$ by infinity)
\begin{equation}f_{nn} = \frac{4(2n-2)!}{T (2^n n!)^2},
\qquad \chi \ll 1.
\end{equation}
Using Stirling's formula
$n!\approx \sqrt{2\pi n}(n/e)^n$, we can write for $n\gg 1$
\begin{equation}
f_{nn} \approx \left(T\sqrt{\pi n^5}\right)^{-1}.
\label{Stir}
\end{equation}
This function corresponds to the QJS (\ref{J-diag})
with
\begin{equation}
F(\hat{n})=F_5(\hat{n}) \equiv (\hat{n}+1)^{-5/4}, \quad
\gamma =\gamma_5 \equiv (T\sqrt{\pi})^{-1}.
\label{V}
\end{equation}
Thus, differently from the case of two-level detector, in the
simplest version of the oscillator detector model the lowering
operator contains the factor $(\hat{n}+1)^{-5/4}$, instead of
$(\hat{n}+1)^{-1/2}$ as in the ``E-model'' (\ref{JE})
 or simply $\hat{1}$ as in the SD model (\ref{JSD}).

The case $\chi\ll 1$ is not very realistic from the practical point
of view, since it corresponds to the detector with very low
efficiency. However, we can calculate the integral (\ref{fmn-osc})
with arbitrary $\xi$ approximately, assuming that $n \gg 1$ and
 using the {\em method of steepest descent\/}.
Rewriting the integrand as $\exp[G(z)]$, one can easily verify
that the points of maxima of the function
\[
G(z)= 2\ln[\sin(z)] +2(n-1)\ln[\cos(z) +\xi\sin(z)] -2\xi n z
\]
are given by the formula
$ z_k = \pm z_0 + k\pi$, where
\be
z_0 =\tan^{-1}(\mu), \quad
\mu = \left(\xi^2 n +n -1\right)^{-1/2},
\ee
 $k=0,1,2,\ldots$ for the plus sign and $k=1,2,\ldots$
for the minus sign.
One can verify that
\be
\exp\left[G(z_k)\right]=
\frac{\mu^2(1+\xi\mu)^{2n-2}}{(1+\mu^2)^n}
\exp\left(-2z_0 \xi n - 2\xi\pi n k\right).
\label{Gk}
\ee
The second derivatives of the function
$G(z)$ at the points of maxima do not depend on $k$:
\be
G''(z_k)= - \frac{4n (\xi^2 +1)}{1+\xi \mu}.
\label{Gpr}
\ee
 Using Eqs. (\ref{Gk}) and (\ref{Gpr})
and performing summation over $k$ we find (taking $Z=\infty$)
\be
f_{nn} = \frac{\chi\sqrt{8\pi} (1+\xi \mu)^{2n-3/2}
\exp\left(-2z_0 \xi n\right)}
{T\sqrt{n} (n +\chi^2 -1) (1+\mu^2)^{n}}
\coth\left(\xi n\pi\right),
\label{fnn-chism0} \ee Although the application of the steepest
descent method can be justified for $n\gg 1$, formula
(\ref{fnn-chism0}) seems to be a good approximation for $n\sim 1$,
too. For example, for $n=1$ (when $\mu=\xi^{-1}$) it yields \be
Tf_{11} \approx 4\chi\sqrt{\pi}\coth(\pi\xi)
\exp\left[-2\xi\tan^{-1}\left(\xi^{-1}\right)\right], \label{appr}
\ee and the numerical values of (\ref{appr}) in the whole interval
$0<\chi<1$ are not very far from the exact value $Tf_{11}=1$,
which holds independently of $\chi$, as far as the upper limit of
integration in (\ref{fmn-osc})
 can be extended to the infinity.

For $n\gg 1$ (when $\mu\ll 1$)
 Eq. (\ref{fnn-chism0}) can be simplified as
\be
f_{nn}(\chi) \approx \frac{\chi\sqrt{8\pi}}{eT}
n^{-3/2}\coth\left(\frac{\chi n\pi}{\sqrt{1-\chi^2}}\right),
\quad \chi \le 1.
\label{fnn-chism}
\ee
For $\chi\ll 1$ the function (\ref{fnn-chism}) assumes the
form (\ref{Stir}), with slightly different coefficient
$\gamma'=(eT)^{-1}\sqrt{8/\pi}
 \approx 1.04 \gamma_5$.

For $\chi>1$ (when parameter $\eta_0$ is imaginary) we have,
instead of (\ref{fmn-osc}),
the integral (considering diagonal coefficients only)
\begin{eqnarray}
f_{nn} &=& \frac{4\chi }{T(\chi^2-1)^{3/2}}
 \int_{0}^{Y} dz\, [\cosh( z) +\zeta \sinh(z)]^{2n-2}
 \nonumber \\ && \times
 \sinh^2( z)
 \exp\left(-2n\zeta z \right),
\label{fmn-osc2}
\end{eqnarray}
where
\be
\zeta={\chi}/{\sqrt{\chi^2 -1}}, \quad
Y= {\lambda T}/(2\zeta).
\ee
Applying again the steepest descent method,
we have now the only point of maximum
\be
z_{max}=\tanh^{-1}(\nu), \quad \nu =
\left[\left(\zeta^2 -1\right)n +1\right]^{-1/2}.
\label{zmax}
\ee
Taking into account the value of the second derivative
of the logarithm of integrand at this point,
\be
G''(z_{max})= -\frac{4n (\zeta^2 -1)} {1 +\zeta\nu},
\quad \zeta\nu = \frac{\chi}{\sqrt{n + \chi^2 -1}},
\label{Gprchibig}
\ee
we obtain
\be
f_{nn}=\frac{\chi\sqrt{8\pi} (1+\zeta \nu)^{2n-3/2}
(1-\nu)^{n(\zeta-1)} }
{T\sqrt{n} (n +\chi^2 -1) (1+\nu)^{n(\zeta+1)}}.
\label{fchi>}
\ee
One can check that the limit of formula (\ref{fchi>})
at $\chi \to 1$ coincides with the analogous limit
of formula (\ref{fnn-chism0}),
so the transition through the point $\chi=1$ is
continuous.

The asymptotical form of (\ref{fchi>})
for $n\gg \chi^2$ is the same as
(\ref{fnn-chism}), except for the last factor:
\be
f_{nn}(\chi) \approx \frac{\chi\sqrt{8\pi}}
{eT} n^{-3/2},
\quad \chi \ge 1,
\label{fnn-chi>1}
\ee
Applying the steepest descent method to the integral
(\ref{int-1}) (for $n\gg 1$), we obtain
the same result (\ref{fnn-chi>1}) with $\chi=1$.
Thus for $\chi \sim 1$ (not too small and not too big)
we obtain the QJS in the form (\ref{J-diag}) with
\be
F(\hat{n}) = F_3(\hat{n}) \equiv (\hat{n}+1)^{-3/4}, \quad
\gamma = \gamma_3 \equiv \frac{\chi\sqrt{8\pi}}{eT}.
\label{F34}
\end{equation}

For very big values of parameter $\chi$
(exceeding $\sqrt{ n} $)
the steepest descent method cannot be used, because
the second derivative of the logarithm of integrand,
given by Eq. (\ref{Gprchibig}), becomes small, and because
the coordinate $z_{max}$, determined by Eq. (\ref{zmax}),
tends to infinity,
while the upper limit $Y$ of integration
in (\ref{fmn-osc2}) tends to the
fixed value $\lambda T/2$.
For $\chi \gg 1$, Eq. (\ref{F34}) holds for the values
of $n$ satisfying approximately the inequality
$n > n_*\sim 4\chi^2 \exp(-\lambda T)$. If $n<n_*$,
then it can be shown that Eq. (\ref{fmn-osc2}) leads
to the same approximate formula (\ref{interp}) as in the
model of two-level detector, so the SD super-operator
(however, without off-diagonal elements)
is restored for relatively not very big values of $n$.

\begin{figure}[h]
\begin{center}
{\includegraphics[height=2.5truein,width=2.5truein]{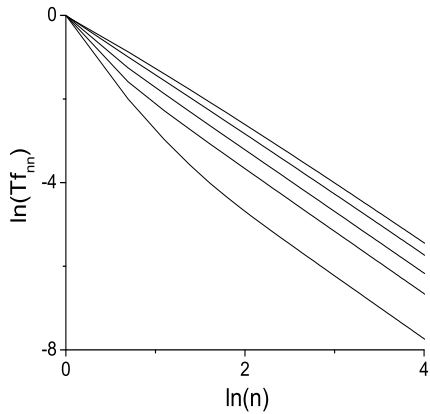}}
\end{center}
\protect\caption{Dependence of diagonal coefficients $f_{nn}$ on the
number $n$, obtained by numerical integration of (\ref{fmn-osc}) and
(\ref{fmn-osc2}) with the fixed value $\lambda T=10$, for small and
moderate values of the parameter $\chi$ (from below):
$\chi=0.1,0.3,0.5,0.8,1.1$.} \label{fig1}
\end{figure}

\begin{figure}[h]
\begin{center}
{\includegraphics[height=2.5truein,width=2.5truein]{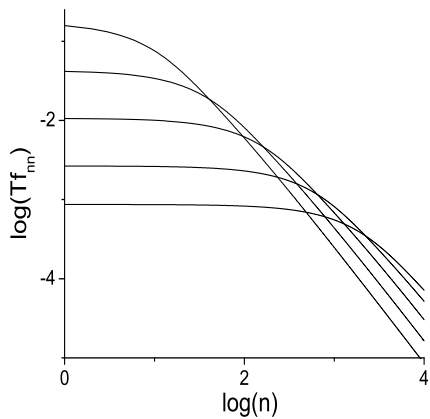}}
\end{center}
\protect\caption{Dependence of diagonal coefficients $f_{nn}$ on
the number $n$, obtained by numerical integration of
(\ref{fmn-osc2}) with the fixed value $\lambda T=10$, for big
values of the parameter $\chi$ (from above): $\chi=5,10,20,40,70$.
Notice the appearance of plateaus corresponding to SD model for
initial values of $n$; for large $n$ they are transformed in
curves with the slope given by power-law dependence.} \label{fig2}
\end{figure}

\begin{figure}[h]
\begin{center}
{\includegraphics[height=2.5truein,width=2.5truein]{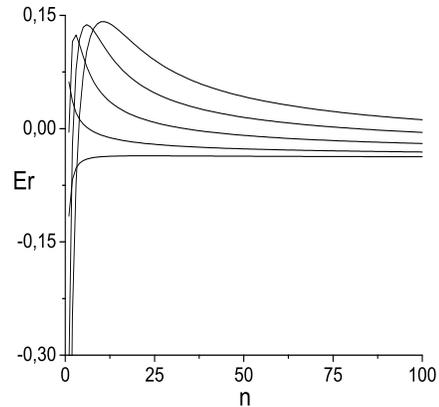}}
\end{center}
\protect\caption{Comparison of numerical integration of
(\ref{fmn-osc}) and (\ref{fmn-osc2}) with the approximate
analytical formulas (\ref{fnn-chism0}) and (\ref{fchi>}) for
$\chi=0.5,1.1,2,3,4$ (from below). We defined relative error by
$Er=(f_{nn}^{num}-f_{nn}^{anal})/f_{nn}^{num}$.} \label{fig3}
\end{figure}

In Figures \ref{fig1} and \ref{fig2} we show the dependence of
diagonal coefficients $f_{nn}$ on the number $n$ for different
values of parameter $\chi$, obtained by numerical integration of
(\ref{fmn-osc}) and (\ref{fmn-osc2}) for the fixed value of the
parameter $\lambda T = 10$; in figure \ref{fig3} we
compare them with the approximate analytical formulas
(\ref{fnn-chism0}) and (\ref{fchi>}). We see that
the coincidence is rather satisfactory for big values of $n$,
although there are some differences for $n\sim 1$.
We also see in Figure \ref{fig2} that the increase of parameter
$\chi$ results in the appearance of the SD plateau for small
values of $n$, which goes into a slope corresponding to the
power-law dependence for big values of $n$.
The height of plateaus diminishes as $\chi^{-2}$ in accordance
with Eq. (\ref{interp}), because big values of $\chi$ correspond
(for fixed values of $\lambda$ and $T$) to small coupling
coefficient $|g|^2$ between the field and MA and, consequently,
low probability of photocount.

\section{Conclusions}

Here we presented two microscopic models for deducting QJS's. In the
first one we supposed that the detector behaves like a 2-level atom,
and in the second -- as a harmonic oscillator. The main difference
between our models and previous ones is that we take into account
the dissipative effects that arise when one couples the actual
detector to the phototube. This scheme includes the spontaneous
decay of the detector with originated photoelectron emission inside
the phototube, which is amplified and viewed as macroscopic electric
current. Using quantum trajectories approach we deduced general
time-dependent transition super-operator, responsible for taking out
a single photon from the field. Since it depends explicitly on
interaction time, we proposed two distinct schemes for obtaining
time independent QJS\'{}s from it. In the first case we assumed that
the interaction time is small and that there are few photons in the
cavity; in this situation we recovered the QJS proposed by Srinivas
and Davies in both detector models. As a second scheme, we
calculated time-averaged QJS on the time interval during which a
photon is certainly absorbed; as the result, we obtained different
non-linear QJS's for the 2-level atom model and the model of
harmonic oscillator. In particular, we have shown that for quantum
states with the predominant contribution of Fock components with big
values of $n$, the QJS has the {\em nonlinear\/} form (\ref{J-diag})
with the power-law asymptotic function
$F(\hat{n})=(\hat{n}+1)^{-\beta}$. However, the concrete value of
the exponent $\beta$ is model-dependent.
 For the 2-level atom model we obtained $\beta=1/2$, whereas
in the model of harmonic oscillator the values $\beta=5/4$ and
$\beta= 3/4$ were found, depending on the ratio between the
spontaneous decay frequency of the excited state and the effective
frequency of coupling between the detector and field mode. Also, we
have demonstrated how the simple Srinivas--Davies QJS arises in the
case of states with small number of photons. Another important
result we obtained is that the QJS's, when applied to density
matrix' non-diagonal elements, are null in average in both models
due to the strong oscillations of the free field terms.

\begin{acknowledgments}
Work supported by FAPESP (SP, Brazil)
contracts \# 00/15084-5, 04/13705-3. SSM and VVD acknowledge
partial financial support from CNPq (DF, Brazil).
\end{acknowledgments}

\end{document}